\documentclass[graybox]{svmult}

\usepackage{graphicx}        
\usepackage{multicol}        
\usepackage[bottom]{footmisc}

\usepackage{hyperref}

\usepackage{newtxtext}       %
\usepackage{newtxmath}       

\usepackage{url} 
\usepackage{ulem}

\usepackage{physics}

\makeindex             

\begin{document}

\title*{Pattern formation in driven condensates}
%

\author{
Kiryang Kwon$^1$ and Jae-yoon Choi$^2$
}

\institute{
$^1$MIT-Harvard Center for Ultracold Atoms, Research Laboratory of Electronics and Department of Physics, Massachusetts Institute of Technology, Cambridge, Massachusetts 02139, USA\\
$^2$Department of Physics, Korea Advanced Institute of Science and Technology, Daejeon 34141, Korea\\
\texttt{Email:jaeyoon.choi@kaist.ac.kr}
}
\maketitle
\abstract
{
Spontaneous pattern formation out of homogeneous media is one of the well-understood examples of hydrodynamic instabilities in classical systems, which naturally leads to the question of its manifestation in quantum fluids. Bose-Einstein condensates (BECs) of atomic gases have been an ideal platform for studying many-body quantum phenomena, such as superfluidity, and simultaneously providing an opportunity to broaden our understanding of classical hydrodynamics into quantum systems. In this review, we introduce a range of experimental studies on the pattern formation in quantum fluids of atomic gases under external driving, including Faraday waves in one and two dimensions, surface patterns, and counterflow instabilities in a mixture of superfluids. The pattern formation in the quantum system can be understood through the parametric amplification process, where an unstable dynamical mode can be exponentially amplified, similar to classical systems. Remarkably, the governing equations for surface excitations of trapped BECs can be mathematically equivalent to those of shallow water, indicating a universal description of the hydrodynamic instability across classical and quantum domains. However, the condensates, as superfluids, also possess fundamental quantum characteristics, such as quantized vorticity and a distinct dissipation channel. These unique features showcase many-body fragmentation under strong modulation and the generation of vortices in the nonlinear regime, which could offer a pathway to the study of quantum turbulence. Furthermore, the coexistence of long-range phase coherence and density modulation in driven condensates could provide unexplored features, such as those seen in supersolid-like sound modes, within nonequilibrium settings. Together with the rapid development of experimental techniques, quantum gas systems can continue to serve as a delicate platform for simulating and understanding complex hydrodynamic phenomena, thereby bridging classical and quantum fluid dynamics.
}

\section{Introduction}

\begin{figure}[b]
\sidecaption[t]
\includegraphics[scale=.8]{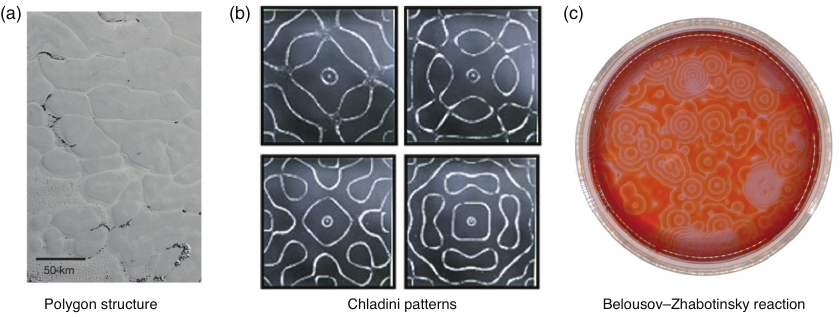}
\caption{
\textbf{Various patterns in classical systems.}
(a) Polygonal structure observed in Sputnik Planum, Pluto.
Adapted with permission from ref~\cite{McKinnon2016}, Springer Nature Limited.
(b) Various Chladni patterns in oscillating square-shaped metal plate. 
Adapted from ref~\cite{Tuan2018} under a Creative Common license \href{https://creativecommons.org/licenses/by/4.0/}{CC BY 4.0}.
(c) oscillatory pattern in a ferroin catalyzed Belousov–Zhabotinsky reaction. 
Adapted from ref~\cite{Howell2021} under a Creative Common license \href{https://creativecommons.org/licenses/by/4.0/}{CC BY 4.0}.
}
\label{fig:Classical}
\end{figure}

Pattern formation is an ubiquitous phenomenon in nature, observed across diverse scales and environments (Fig.~\ref{fig:Classical}), from ripples in a kitchen sink to the Belousov-Zhabotinsky chemical reaction~\cite{Cross1993}, the intricate morphological development of living organisms~\cite{Koch1994}, and even Pluto's mysterious polygonal structure~\cite{McKinnon2016}. In classical systems, pattern formation has been understood from linear stability analysis of nonlinear amplitude equations, which explains how perturbations grow with spatial structures and how they become saturated by nonlinear effects~\cite{Cross1993,Koch1994}. The initial stage of pattern formation, where non-uniform patterns spontaneously emerge from a homogeneous state, has been well understood through linear instability analysis. For a system near linear instability, its dynamics can be accurately described by nonlinear amplitude equations~\cite{Swift1977}. These equations can successfully explain the key mechanisms of pattern formation, such as spatial structure growth from disturbances, saturation of the growth due to nonlinear effects, interplay between substances of different diffusion rate, and, in the case of dynamical instability, wave propagation and dispersion~\cite{Cross1993,Koch1994}. Despite these understandings, the fundamental question of how patterns spontaneously emerge from a uniform system remains a challenge, especially beyond the linear stability regime. 

Bose-Einstein condensates (BECs) of atomic gases have been considered as an analog quantum simulator that could solve complex quantum many-body problems, providing a unique opportunity for the study of nonequilibrium spontaneous pattern formation~\cite{Bloch2012,Langen2015}. Ultracold atomic gases have no defects or impurities and offer unprecedented high controllability in experimental parameters, such as interaction strength by using Feshbach resonances~\cite{Chin2010}. Moreover, over the last decades, the imaging and the potential engineering techniques have been largely improved, facilitating quantitative comparison with theoretical models. For example, three-dimensional spin vector configurations of one-dimensional quantum gases can be measured by spin-resolved imaging techniques~\cite{Kunkel2019} and an engineered optical potential can remove the harmonic confinement from trapping laser, confining atoms in a box potential with homogeneous density distribution~\cite{Navon2021}. Leveraging these technical advances, the atomic quantum simulator has explored various aspects of nonequilibrium quantum dynamics: universal behavior in far from equilibrium~\cite{Navon2016,Prufer2018,Erne2018,Glidden2021,Dogra2023,Huh2024}, defect formation across phase transitions~\cite{Lamporesi2013,Corman2014,Navon2015,Ko2019}, exotic phases of matter in nonequilibrium~\cite{Schreiber2015,Choi2016}, as well as pattern formation, such as Faraday waves in one dimension~\cite{Engels2007,Smits2018,Nguyen2019,Hernandez2021}, and controlled patterns in two dimensions~\cite{Clark2017,Fu2018,Kwon2021,Liebster2025}. These studies showcase phenomena analogous to classical hydrodynamics while simultaneously extending our understanding into the distinct realm of quantum fluid dynamics. In this chapter, we review the experimental progress on pattern formation of quantum gases and explore how the classical intuition of instabilities manifests in the collective dynamics of quantum fluids.

\section{Faraday waves in superfluids}
\label{sec:2}
Superfluidity is one of the most intriguing phenomena in nature, characterized by zero-resistance flow and zero viscosity~\cite{Kapitza1938}, which marks a stark contrast to classical fluids. Atomic gases in the few hundred nK regime can become Bose-Einstein condensates or degenerate Fermi gases and can exhibit superfluidity~\cite{Raman1999,Zwierlein2005}. In particular, quantum gas systems offer various experimental toolkits that manipulate and observe the quantum state of the system, representing an ideal platform for studying novel features of quantum fluids. In this chapter, we review experimental studies of the Faraday pattern, observed in the surface of a vertically oscillating classical fluid~\cite{Faraday1831}, in both bosonic and fermionic atomic systems. Various approaches have been employed to observe the Faraday pattern, including trap frequency modulation~\cite{Engels2007,Smits2018}, interaction modulation via Feshbach resonances~\cite{Clark2017,Fu2018,Nguyen2019,Zhang2020}, and tailoring the external potential~\cite{Liebster2025}.
These include early observations in BECs, extensions to two-dimensional traps~\cite{Kwon2021,Liebster2025}, reinterpretations within the framework of time crystals~\cite{Smits2018}, and recent realizations in strongly interacting Fermi gases~\cite{Hernandez2021}.

\subsection{The first observation of Faraday waves in quantum fluids}

\begin{figure}[t]
\sidecaption[t]
\includegraphics[scale=.65]{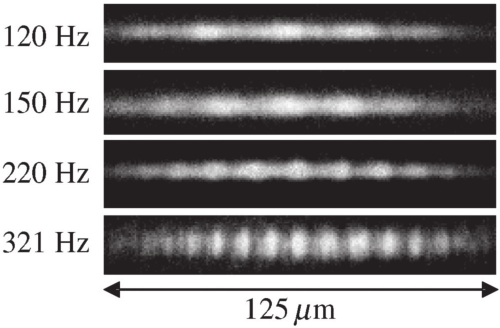}
\caption{\textbf{Faraday pattern in Bose-Einstein condensates.} Absorption images of trapped quasi one-dimensional BEC after modulating the radial trap frequency. Labels in each image represent the driving frequencies. Adapted with permission from~\cite{Engels2007}, APS.
}
\label{fig:FW}
\end{figure}

The first observation of Faraday waves in an ultracold atomic gas was pioneered by Peter Engels and his colleagues~\cite{Engels2007}. 
In this experiment, a cigar-shaped $^{87}$Rb Bose-Einstein condensate (BEC), containing approximately $5\times10^5$ atoms, was prepared in a Ioffe-Pritchard magnetic trap~\footnote{The Ioffe-Pritchard magnetic trap is a three-dimensional magnetic trap that has cylindrical symmetry. Because the magnetic field at the trap center is nonzero, it prevents non-adiabatic spin-flip atom losses (Majorana loss).
}.
It provided a harmonic confinement for the atoms with the radial and axial trapping frequencies of $\{\omega_{r}/2\pi,\omega_z/2\pi\}=\{160.5,7\}$~Hz.
After periodic modulation of the current for the magnetic trap, the density of the condensate oscillated periodically, which resulted in the formation of the Faraday pattern.
Figure~\ref{fig:FW} displays the observed atomic density distribution of the BEC after 10 to 30 oscillations at different modulation frequencies. 
It breaks the continuous symmetry typically seen in a harmonically trapped BEC, where density modulations with shorter characteristic wavelength was observed for higher driving frequencies. 

\label{subsec:2.1}

It is notable that this experiment utilized the radial breathing mode to observe the Faraday pattern instead of modulating the atomic scattering length~\cite{Staliunas2002,Kevrekidis2004}. In this experiment, the radial breathing motion modulates the atomic density and thereby the nonlinear interactions of the condensates, leading to a longitudinal density modulation. Even at the early stage of the study, the experiment revealed many interesting features, which inspired many other research groups. For instance, the study demonstrated that under strong modulation, the condensates behave like an impact oscillator. This led to the question of how the breakdown of Faraday patterns manifests dynamically.

The experimental protocol developed in this work was later employed and refined to investigate Faraday waves in quasi-one-dimensional condensates.
With the addition of fast, non-destructive imaging, Faraday patterns were reinterpreted as a form of time crystalline order, which we later introduce further in Chapter 2.3. Cominotti and colleagues at Trento extended this protocol to a two-component condensate, where Faraday waves appeared not only as density waves but also spin waves~\cite{Cominotti2022}. Both density and spin waves exhibited massless linear dispersion relations, whereas introducing coherent coupling between the two components with a microwave field gave the spin-wave excitation a gap and a massive dispersion relation, with the density-wave dispersion remaining unaffected.

\subsection{From Faraday waves to granulation}
\label{subsec:2.2}
\begin{figure}[t]
\sidecaption[t]
\includegraphics[scale=.65]{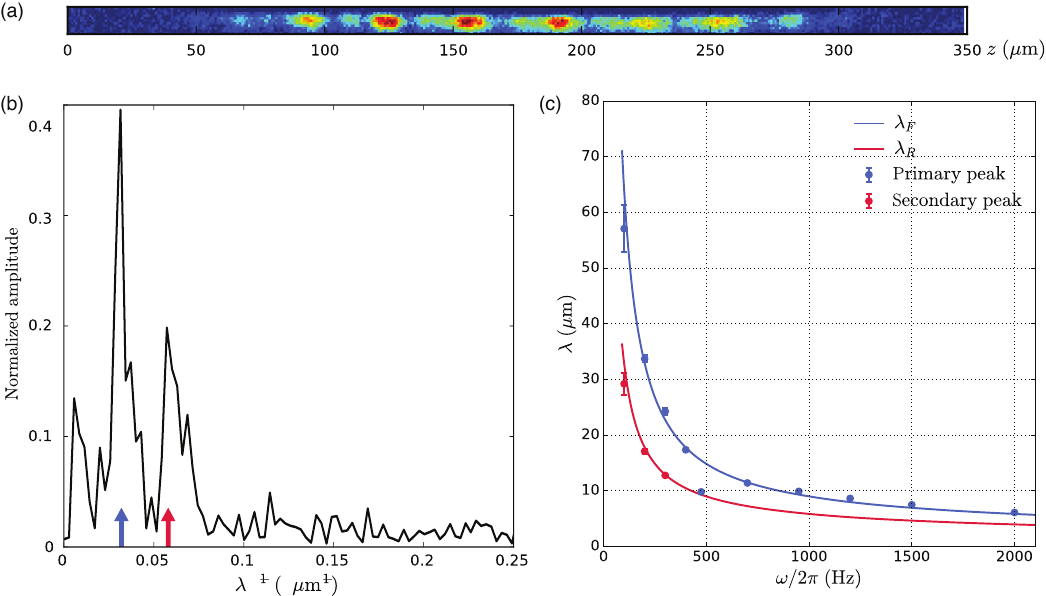}
\caption{
\textbf{Faraday and resonant waves in one-dimensional Bose gases.}
(a) Absorption images of trapped Bose-Einstein condensate after modulating the scattering length at $\omega/2\pi=200$~Hz. (b) Fast-Fourier transform (FFT) spectrum of the line density. The spectrum shows two peaks, where the primary peak (marked by blue arrow) represents the Faraday waves and the second peak (marked by red arrow) indicate the resonant wave. (c) Modulation wavelength dependence on the driving frequency $\omega$. The solid lines are the wave lengths obtained from a 3D variational calculation~\cite{Nicolin2011} for the Faraday waves (blue line) and resonance wave (red line), respectively.  Labels in each image represent the driving frequencies. 
Adapted from~\cite{Nguyen2019} under the terms of the \href{https://creativecommons.org/licenses/by/4.0/}{CC BY 4.0}.
}
\label{fig:FW2}
\end{figure}

The impact of strong modulation on superfluids has been re-investigated by Nguyen and colleagues at Rice University~\cite{Nguyen2019}.  They conducted a series of experiments on parametric excitation in Bose–Einstein condensates (BECs) using ultracold $^7$Li atoms. The key ingredient of this study is the tunability of $s$-wave scattering length by using the Feshbach resonance~\cite{Chin2010}. In particular, the $^7$Li atom has a broad Feshbach resonance with moderate atom loss~\cite{Pollack2009}, allowing them to modulate the interaction strength directly. Moreover, the atoms are confined in an one-dimensional elongated optical dipole trap whose harmonic frequencies are $\{\omega_{r}/2\pi,\omega_z/2\pi\}=\{475,7\}$~Hz, and the modulation of the trapping potential from the magnetic field modulation is negligible. This method resulted in significantly reduced atom loss under strong modulation compared to the trap modulation, enabling more precise measurements and longer duration times.

In this study, they also observed Faraday waves after modulating the scattering length, which is only clearly seen at higher driving frequencies. In the high frequency regime ($\omega/2\pi > 100$~Hz), the atomic density displayed spatial modulations at two distinctive wavelengths, $\lambda_R$ and $\lambda_F$ [Fig.~\ref{fig:FW2}(a) and (b)]. These two waves, each called resonant waves and Faraday waves, both arise from parametric instabilities in nonlinear systems. Resonant waves are characterized by twice faster oscillation and shorter wavelengths than Faraday waves~\cite{Nicolin2011}. Figure~\ref{fig:FW2}(c) presents the frequency dependence of the density modulation wavelength, showing excellent agreement with the 3D variational calculation~\cite{Nicolin2011} and elucidating the effect of the trapping potential under periodic driving.

\begin{figure}[t]
\sidecaption[t]
\includegraphics[scale=.65]{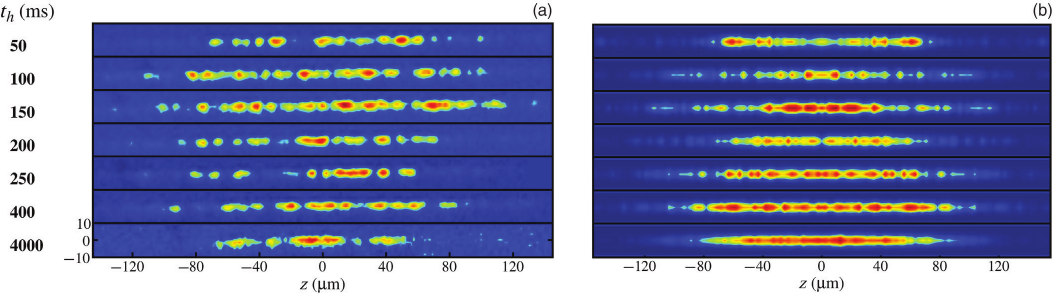}
\caption{
\textbf{Granulation of a Bose-Einstein condensate,}
(a) Experimental images and (b) Gross-Pitaevskii (GP) simulations of density profiles at different modulation times $t_h$ with low modulation frequency $\omega/2\pi=70$~Hz. 
Adapted with permission from ref~\cite{Nguyen2019} under a Creative Commons license \href{https://creativecommons.org/licenses/by/4.0/}{CC BY 4.0}.
}
\label{fig:GR}
\end{figure}

Remarkably, at lower modulation frequencies, they did not observe regular Faraday patterns. To find any particular pattern formation, they increased the modulation strength while keeping their condensate fraction high. Instead, they found irregular, fragmented density distributions (Fig.~\ref{fig:GR}), a phenomenon referred to as granulation. The term is borrowed from the respective classical state composed of spatially inhomogeneous discrete macroscopic particles~\cite{Jaeger1996}. This granulation of a BEC had not been previously predicted by mean-field simulations. To capture this phenomenon they used a Multiconfigurational Time-Dependent Hartree theory for Bosons (MCTDHB)~\cite{Streltsov2007,Alon2008}, an ab-initio many-body approach that goes beyond the mean-field description. The MCTDHB method allows for the inclusion of quantum correlations and fragmentation by representing the many-body wavefunction as a linear combination of time-dependent single-particle functions. Studying the density-density correlation function of the fragmented condensate, they found that the granulated state was accompanied by the emergence of higher-order eigenvalues of the reduced one-body density matrix, indicating the granulation is purely driven by a quantum many-body effect.

In a follow-up theoretical work, the mechanism for the granulation was analyzed in detail~\cite{Lode2021}. It was shown that interaction modulation first induces phase modulations in the condensate wavefunction, which subsequently evolve into density modulations. Below a certain threshold frequency, the condensate responds quasi-adiabatically to the external drive, maintaining coherence. However, as the modulation frequency increases, the system accumulates a large number of excited modes, which leads to irregular phase dynamics and generates nonlinear interference among multiple density modes. Ultimately, the system becomes a fragmented many-body state.

\subsection{Time crystals}
\label{subsec:3.3}

\begin{figure}[b]
\sidecaption[t]
\includegraphics[scale=.8]{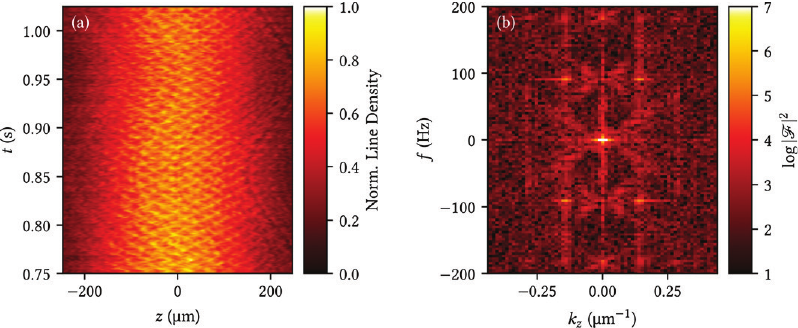}
\caption{
\textbf{Long-term stability of space-time crystal.}
(a) Phase contrast images of a single condensate driven at $f_D$=183.2~Hz recorded within 1.1ms time interval. (b)  Two-dimensional Fourier transform in space and time in logarithmic scale. It reveals a spatial period of $k_z=0.14 \rm{\mu m}^{-1}$ and a temporal period at half the driving frequency $f=91.6$~Hz.
Adapted from~\cite{Smits2020} under the terms of the \href{https://creativecommons.org/licenses/by/4.0/}{CC BY 4.0}.
}
\label{fig:TC}
\end{figure}

Recent studies have proposed that Faraday pattern formation in Bose–Einstein condensates can be interpreted as a manifestation of novel nonequilibrium phases of matter. One such phase is the time crystal. Originally proposed by Wilczek~\cite{Wilczek2012}, it refers to spontaneous symmetry breaking in time translational symmetry, resulting in a system that exhibits periodic motion in the ground state. Following Wilczek’s proposal, a no-go theorem was established~\cite{Watanabe2015}, showing that continuous time-translation symmetry cannot be broken in equilibrium ground states. This shifted the focus towards Floquet time crystals, which emerge under periodic driving and exhibit subharmonic responses, i.e., oscillations at integer multiples of the driving period. Several experiments have since demonstrated such behavior in trapped ions~\cite{Zhang2017}, nitrogen-vacancy centers~\cite{Choi2017}, nuclear magnetic resonance (NMR) systems~\cite{Pal2018,Rovny2018}, and superfluid $^3$He~\cite{Autti2018}.

Faraday waves, exhibiting a subharmonic response to periodic driving, emerge as a natural candidate for a time crystal. While spontaneous breaking of both the spatial and temporal symmetry makes them especially intriguing, the robustness of Faraday waves as a time-crystalline phase has remained an open question. In this context, Smits and colleagues at Utrecht University~\cite{Smits2018,Smits2020} investigated time-crystalline behavior in a quasi-one-dimensional Bose–Einstein condensate with trap frequencies of $\{\omega_{r}/2\pi,\omega_z/2\pi\}=\{92,5\}$~Hz. By periodically modulating the radial trapping frequency, they excited Faraday waves along the axial direction.Using phase contrast imaging, they were able to record up to 250 consecutive images of the same condensate with minimal disturbance, enabling a high-resolution space-time analysis of the density pattern.

Their observations revealed that the emerging density modulation oscillates at exactly half the frequency of the external drive, a hallmark of discrete time-translation symmetry breaking~\cite{Sacha2018}. Moreover, the pattern remained phase-coherent over many oscillation cycles, indicating long-term stability. The authors modeled this by introducing a phenomenological fourth-order interaction term in an effective Hamiltonian for the excited mode. They demonstrated that this nonlinear term plays a key role in saturating the amplitude of the mode, thereby stabilizing the emergent space-time crystal against runaway growth.

This work provides a compelling reinterpretation of Faraday pattern formation as a realization of a Floquet space-time crystal and highlights how coherent nonlinear dynamics in clean quantum systems can give rise to stable, self-organized temporal structures far from equilibrium.

\subsection{Faraday waves in fermionic gases}
\label{subsec:2.4}
Lastly, we would like to introduce the experimental study of the Faraday wave in fermionic gases from Hernández-Rajkov and colleagues~\cite{Hernandez2021}. It has been numerically shown that the Faraday wave can be observed in fermionic superfluid across the BCS-BEC crossover~\cite{Capuzzi2008,Tang2011}, which highlights the universal description of the parametric instability in quantum fluids. In this study, ultracold $^6$Li fermionic atoms were prepared in an elongated one-dimensional optical dipole trap with frequencies $(\omega_r,\omega_z)=2\pi\times(163, 11)$~Hz. To study pattern formation, the laser intensity for the dipole trap was modulated while the interaction strength remained constant.

\begin{figure}[t]
\sidecaption[t]
\includegraphics[scale=.83]{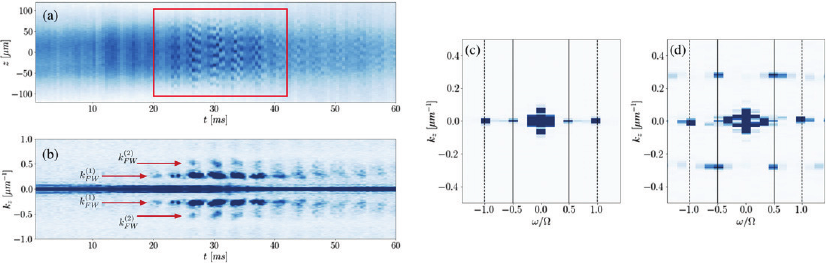}
\caption{
\textbf{Faraday waves and space-time crystal in ultracold one-dimensional Fermi gases.}
(a) Integrated optical density and (b) its Fourier transformation under periodic modulation. Rectangle region in (a) highlights the formation of Faraday waves in a molecular BEC. (c) Space-time Fourier transformation in the early stage ($t<20$~ms) and (d) after the onset of the Faraday pattern ($20<t<40$~ms). 
The driving frequency $\omega$ is represented in the units of radial trap frequency $\Omega$.
Dashed lines represent the breathing mode frequency, and  solid lines indicate the Faraday wave frequency, respectively. 
Adapted from ref~\cite{Hernandez2021} under the terms of the \href{https://creativecommons.org/licenses/by/4.0/}{CC BY 4.0}.
}
\label{fig:TC}
\end{figure}

In this work, the system was tuned to the BEC side of the Feshbach resonance, where fermionic atoms form tightly bound pairs. They observed the characteristic feature of Faraday patterns previously realized in Bose gases~\cite{Engels2007,Smits2018,Nguyen2019}, where the density modulation oscillated at half the frequency of the external drive. Moreover, by analyzing the wavevector and frequency of the observed Faraday wave, they extracted a phase velocity that was found to be in good agreement with the speed of sound in the Fermi superfluid. This suggests that Faraday waves could serve as alternative probes for sound velocity in fermionic systems. On the BCS side or near the unitary regime, however, they could not observe density modulation. This could be attributed to the low condensate fraction in the strongly interacting regime. Even at zero temperature ($T=0)$, the maximum condensate fraction is about $50\%$~\cite{Giorgini2008}. Because non-condensed atoms reduce the contrast of the density modulation, it is difficult to identify the Faraday wave, just as quantum vortices are hard to identify in the BCS side~\cite{Zwierlein2005}. 

\section{Pattern formation in two dimensions}
\label{sec:4}

While Faraday waves in one dimension are constrained to stripes and fringes, their two-dimensional (2D) counterparts host much richer patterns, such as a triangle, a square and a hexagon geometric shapes. Moreover, in two-dimensional condensates, excitations can propagate in multiple directions, allowing nonlinear interference and mode competition to play a crucial role. In this chapter, we explore recent experimental realizations of pattern formation in 2D, ranging from matter-wave jets~\cite{Clark2017} and regular lattice structures~\cite{Liebster2025} to polygonal surface modes~\cite{Kwon2021}. These experiments illustrate how dimensionality enriches the dynamical response of condensates and provides new tools for probing the collective behavior of quantum fluids.

\subsection{Patterns in weakly interacting driven systems}
\label{subsec:3.1}
\begin{figure}[t]
\centering\includegraphics[scale=.7]{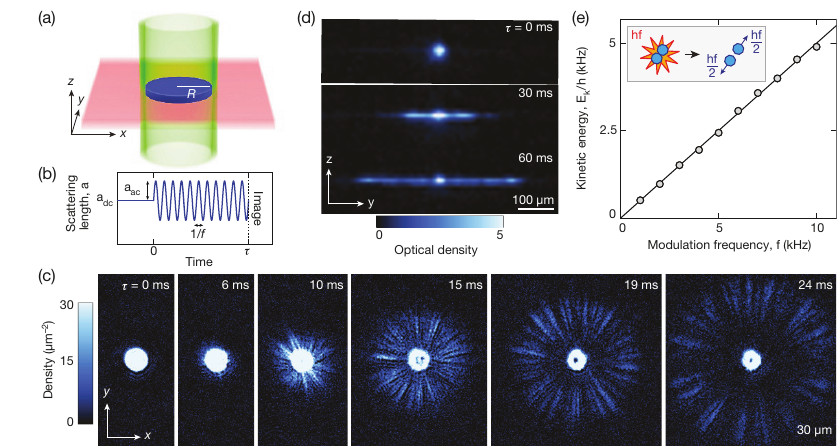}
\caption{
\textbf{Emission of matter-wave jets from Bose-Einstein condensates.}
(a) Bose-Einstein condensate of $^{133}$Cs atoms are confined in a box potential of radius $R$. (b) Modulating the scattering length $a_s(t)=a_{\rm dc}+a_{\rm sc}\sin\omega t$, where $a_{\rm dc}=5~a_B$ and $a_{\rm sc}=60~a_B$. (c) Top down images of the condensate after the modulation at $\omega/2\pi=3.5$~kHz. (d) Side view images of the condensate, displaying emission of the matter-wave in the horizontal plane. (e) Kinetic energy dependence of the modulation frequency $f=\omega/2\pi$. Inset shows a schematic process of jets: two atoms collide, absorb energy $\hbar f$ to produce a pair of moving atoms in opposite directions with kinetic energy $\hbar f/2$. 
Adapted with permission from ref~\cite{Clark2017}, Springer Nature Limited.
}
\label{fig:Firework}  
\end{figure}

An extension of the pattern formation in two-dimensional superfluids is pioneered by Cheng Chin’s group at the University of Chicago~\cite{Clark2017,Fu2018,Zhang2020}. In their first work, called ``Bose firework", they prepared a Bose-Einstein condensate of $^{133}$Cs atoms in a two-dimensional box potential, where the scattering length of the atoms was modulated by using the Feshbach resonance [Fig.~\ref{fig:Firework}(a) and (b)]. Here, the background scattering length $a_{\rm dc}=5a_B$ was set to be much smaller than the modulation amplitude $a_{\rm sc}$, which is essential in this experiment. Unlike a one-dimensional system, the Bogoliubov quasi-particles can have a momentum in many different directions. In two dimensions, thus, one might expect an irregular density pattern after a periodic modulation. The experiments done by the Chicago group found more than that. They observed emission of matter-wave jets that propagated out of the box potential [Fig.~\ref{fig:Firework}(c) and (d)], which resembles a firework explosion. 

Monitoring the distance of the emitted atoms over time, they measured the kinetic energy of the jets $E_k$ and found that it is half of the oscillating field energy, $E_k=\hbar\omega/2$ [Fig.~\ref{fig:Firework}(e)]. This indicates that two atoms absorbed the modulation energy of $\hbar\omega/2$ through a collision, and then ejected in opposite directions with equal energy as a result of the energy and momentum conservation laws. When the momentum-correlated atom pairs moved through the condensate, the atom pairs stimulated further collisions in the same direction, causing macroscopic occupation of the same momentum state and forming matter-wave jets. Since the pair production process is purely quantum mechanical and random, the angular distribution of jets was different between experimental runs. This phenomenon can occur only when the driving amplitude of the modulation $a_{\rm ac}$ exceeds a certain threshold, which is analogous to the light emission in a laser. When the number of atom pairs was insufficient, for example, leaving the trap before an additional collision, the stimulated collision process was suppressed, and they observed diffusive clouds. 
This explains why having a small scattering length, $a_{\rm dc}$, is crucial for observing jet emission. The stimulated collision can also be understood as a parametric amplification process, where created atom pairs produce an effective grating that resonantly amplifies their patterns. Indeed, they showed that irregular density patterns were amplified before the jet emission~\cite{Fu2018}. This work has been recently extended into a spinor Bose condensate of Li$^7$ atoms, a system with spin degrees of freedom~\cite {Kim2021}, demonstrating emission of spin-momentum correlated matter wave jets via dynamic instability in the spin state $|F=1,m_F=0\rangle$.

\begin{figure}[t]
\centering\includegraphics[scale=.65]{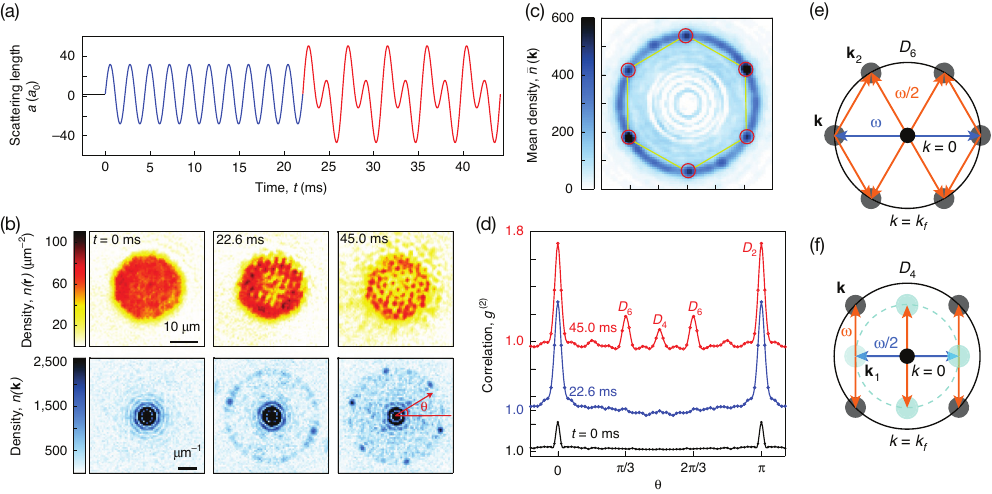}
\caption{
\textbf{Pattern formation in BECs featuring a two-frequency modulation.}
(a) Two stage modulation of the scattering length that generates triangular density waves.
(b) In situ images of the condensate during the modulation (top row) and the corresponding Fourier transform images (bottom row). (c) Averaged Fourier transform image after $t=45$~ms driving. The authors developed a pattern recognition algorithm to align the randomly oriented patterns. 
(d) Angular distribution of a density correlation function. 
(e) Scattering process during the two stage modulation. Top row represents the spontaneous formation of atom pairs with momentum $\mathbf{k}$ of magnitude $k_f$. (f) Bottom row indicates additional pair creation process, where atoms with momentum $\mathbf{k}$ collide with the condensate. 
Adapted with permission from ref~\cite{Zhang2020}, Springer Nature Limited.
}
\label{fig:control}  
\end{figure}

In their follow-up study, they further utilize the stimulated collisional process to control the symmetry of the density wave patterns under periodic driving~\cite{Zhang2020}. The key idea was using a two-stage frequency modulation with two different oscillating frequencies [Fig.~\ref{fig:control}(a)]. For example, a triangular lattice pattern with $D_6$ symmetry was observed after the condensate was driven at $\omega_1=\omega$ and then subjected to a second stage modulation by superimposing $\omega_2=\omega/2$ [Fig.~\ref{fig:control} (b) and (c)]. The density distribution after a single frequency modulation (Fig.~\ref{fig:control}(b) and $t=22.6$~ms) shows irregular patterns because  atom pairs are excited with momentum in random directions Fig.~\ref{fig:control}(c). After the second stage modulation ($t=45$~ms), density waves with $D_6$ symmetry emerged [Fig.~\ref{fig:control}(d)]. 

The symmetric pattern formation can be understood from the wave-mixing process associated with the two-step modulation. Under a single frequency driving, atom pairs with energy $\hbar\omega/2$ and momentum $\pm\mathbf{k}_1$ are generated from the condensate. When the additional frequency $\hbar\omega/2$ was introduced, the atoms with momentum $\mathbf{k}_1$, for instance, can scatter with the condensate and produce atom pairs with momentum $\mathbf{k}$ and  $\mathbf{k}_1-\mathbf{k}$. Because of the energy conservation law, we have 

\begin{equation}
\varepsilon_{\mathbf{k}_1} + \hbar\omega/2 = \varepsilon_{\mathbf{k}}+\varepsilon_{\mathbf{k}-\mathbf{k}_1},
\end{equation}
where left- and right-hand sides are energies before and after the collision, and $\varepsilon_{\mathbf{k}}$ is the kinetic energy for the momentum $\mathbf{k}$. Given that $\varepsilon_{\mathbf{k}_1}=\hbar\omega/2$, this equation can be satisfied if $|\mathbf{k}_1|=|\mathbf{k}|=|\mathbf{k}_1-\mathbf{k}|$ and $\varepsilon_{\mathbf{k}}=\varepsilon_{\mathbf{k}_1-\mathbf{k}}=\hbar\omega/2$. Modulations at $\omega$ followed by $\omega/2$ therefore produce patterns with $D_6$ symmetry [Fig.~\ref{fig:control}(e)], and modulations in reverse order, $\omega/2$ followed by $\omega$, produce patterns with $D_4$ symmetry [Fig.~\ref{fig:control}(f)].

\subsection{Stable Faraday patterns in two dimensions}
\label{subsec:3.2}

\begin{figure}[b]
\centering\includegraphics[scale=.6]{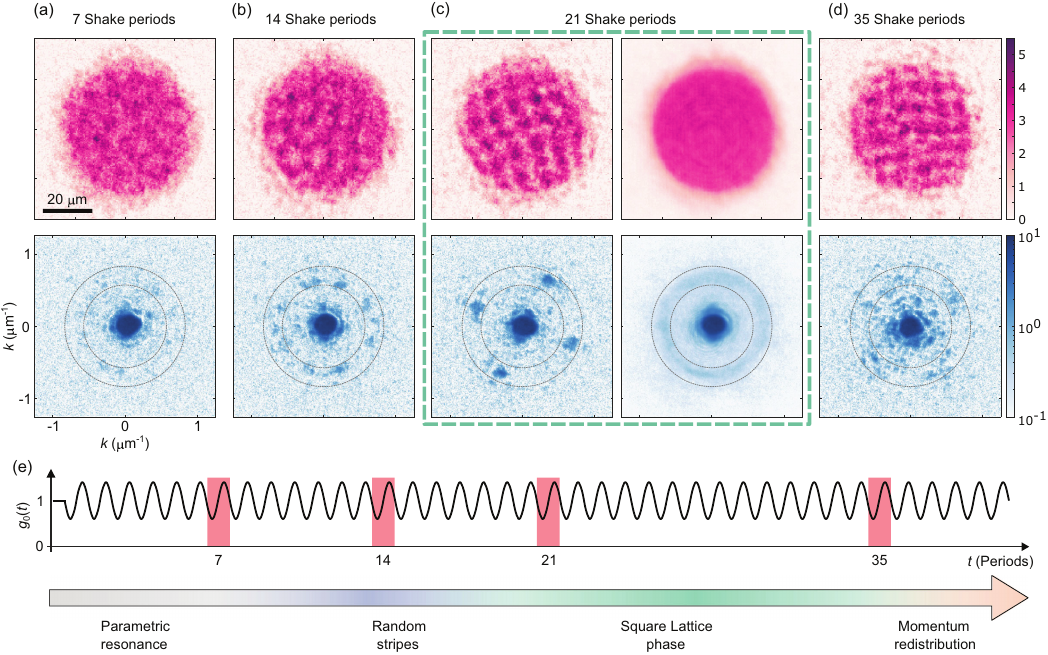}
\caption{
\textbf{Faraday wave in two-dimensional superfluids.}
 (a) Real space (top row) and momentum space (bottom row) images of atomic gases after 7 periods of oscillation at $\omega/2\pi=400$~Hz. The dashed line in the momentum space image indicates the region of resonant momentum $k_c$. (b) After 14 periods, density modulation becomes apparent, and back-to-back correlated peaks also appear in momentum space. (c) A square lattice structure is created after 21 periods. Images in the left column represent distributions of a single experimental run, whereas right column presents averaged distributions. (d) Late time images show a distorted square lattice with other momentum components being populated (e) Periodic oscillation of the interaction strength, and various stages are indicated by the colored arrow. 
Adapted with permission from ref~\cite{Liebster2025} under a Creative Commons license \href{https://creativecommons.org/licenses/by/4.0/}{CC BY 4.0}.
}
\label{fig:FW2D}  
\end{figure}

Can systems with non-negligible interaction have stable Faraday patterns in 2D? In a finite system, this could be non-trivial as the waves reflected from the boundary of a box potential could interfere and destabilize the patterns. Recently, Liebster and colleagues at Heidelberg University developed a novel technique to overcome this difficulty and observed stable Faraday patterns in two-dimensional Bose-Einstein condensates of atomic gases~\cite{Liebster2025}.

In this experiment, a quasi-2D condensate of $^{39}$K atoms was prepared in a uniform trap, but its wall was engineered to have a gradual slope, in contrast to a steep wall of a conventional box trap. The slanted wall acts as a soft boundary, such that density waves generated inside the wall propagate to the boundary where their speed gently slows down because of the gradual reduction in atomic density. This absorbing wall effectively mimics the dynamics in an infinitely extended system. To excite the Faraday wave, they modulated the scattering length via Feshbach resonance, $a_s(t)=\bar{a}_s(1-r \sin\omega t)$, where $\bar{a}_s=100a_B$ is the offset scattering length, $a_B$ is the Bohr radius, modulation amplitude is $|r|<1$, and $\omega$ is the modulation frequency.

Figure~\ref{fig:FW2D} displays representative images of the atomic distribution both in real and momentum space, where the momentum space distribution is obtained by a momentum focusing technique~\cite{Lamporesi2010}. In this method, the atoms expand inside a weak harmonic trap. After a quarter of the trap oscillation period, their \textit{intrap} velocities are mapped onto spatial positions, so that the final atomic distribution directly reflects the intrap momentum distribution. In the early time [Fig.~\ref{fig:FW2D}(a) and (b)], the real space density profiles have no clear patterns. In contrast, in the momentum space, back-to-back momentum-correlated pairs are observed around the characteristic momentum $\hbar k_c$. The underlying mechanism for these correlated pairs is similar to the previous Faraday wave experiments~\cite{Engels2007,Zhang2020}. The external driving creates correlated particle pairs with opposite momentum, where each particle carries half of the energy $\hbar\omega/2$. The characteristic wavelength $k_c$ is determined by the Bogoliubov energy-momentum dispersion. After 21 periods of driving [Fig.~\ref{fig:FW2D}(c)], a square lattice structure is developed both in real and momentum space. The orientation of the lattice is random, confirmed by the uniform density in averaged images, indicating spontaneous symmetry breaking. 
Many other momenta become excited at later times, and the square lattice structure are deformed [Fig.~\ref{fig:FW2D}(d)]. Different phases during the time evolution is shown in Fig.~\ref{fig:FW2D}(e).

Considering the momentum-pair production process is random, which is also shown in Fig.~\ref{fig:FW2D}(b), the formation of a square lattice structure is quite remarkable. They attributed the stable pattern formation to the interplay between parametric amplification and nonlinear mode competition~\cite{Fujii2024}. After momentum-correlated pairs in many different directions are generated, they compete with each other and can suppress the growth of the other mode. In particular, modes at an angle of $\pi/2$ can reinforce the growth of each other through nonlinear interaction. This mutual reinforcement can stabilize the two orthogonal momentum modes and thereby form a stationary square pattern. It is further supported by constructing coupled amplitude equations for two momentum modes, whose solution displays the flow of the amplitude growth trajectories in phase space.They identified a stable fixed point in the dynamics that can be found when the intersection angle is $\pi/2$~\cite{Liebster2025,Fujii2024}. 

\subsection{Surface pattern--formation in driven condensates}
\label{subsec:3.3}

\begin{figure}[b]
\centering\includegraphics[scale=.6]{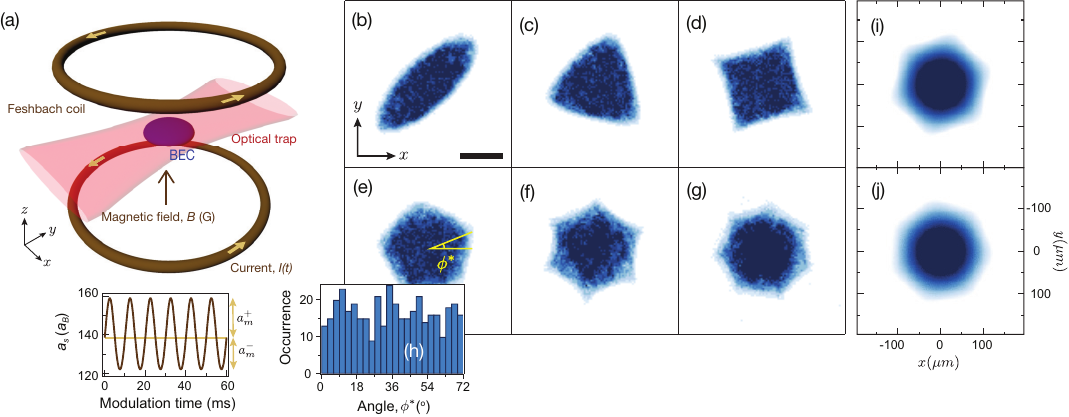}
\caption{
\textbf{Star-shaped patterns in a BEC.} (a) Schematic diagram of the experimental setup. A BEC of $^7$Li atoms is trapped in a hybrid potential. An optical potential (Feshbach field) provided a harmonic confinement along vertical (radial) direction.
Inset below represents the scattering length driven by current modulation of the Feshbach coil. 
(b)-(g) Intrap density distribution (single image) of the atoms under periodic driving. 
Regular polygons from $D_2$ symmetry to $D_7$ symmetry were observed.
(h) Histogram of the orientation angle $\phi^{*}$ for pentagon-shaped BECs, defined in (e). 
Numerical simulations of the (i) hexagon and (j) heptagon shaped patterns by solving the Gross-Pitaevskii equation.
Adapted with permission from ref~\cite{Kwon2021}, American Physical Society.
}
\label{fig:star}  
\end{figure}

So far, we have introduced density waves inside the condensate, caused by instabilities of the bulk. In addition, driven condensates can also exhibit rich surface mode instabilities. One of the well-known examples in classical systems is the Leidenfrost effect~\cite{Leidenfrost1966}, a star-shaped pattern that emerges on the surface of a levitating water droplet above a hot plate~\cite{Ma2017}. This is a result of a surface-mode instability driven by capillary waves excitation. A recent experiment from Kwon and colleagues demonstrated similar behavior in atomic superfluids~\cite{Kwon2021}. In the experiment, $^7$Li condensates were trapped in a hybrid potential made of optical and magnetic traps [Fig.~\ref{fig:star}(a)]. The magnetic confinement is produced by a Feshbach field that provides a symmetric radial confinement with harmonic frequency $\omega_r/2\pi=29.4$~Hz, and an optical trap provides vertical confinement with trap frequency $\omega/2\pi=725$~Hz.

Figure~\ref{fig:star} displays in situ images of regular polygon-shaped surface modes after periodic driving of the scattering length. The orientation of the surface patterns, which spans from quadrupole ($D_2$) to heptagon ($D_7$) symmetries, appeared in random directions, indicating a spontaneous symmetry-breaking process. It was found that resonant frequencies for each $l$-fold symmetry pattern correspond to twice the natural frequencies of surface excitations, $\omega_{\rm res}=2\omega_l=2\sqrt{l}\omega_{r}$. This can be understood by parametric resonance of the surface mode~\cite{Maity2020}, where the disturbance of the density follows the equation for a parametrically driven oscillator with a natural frequency $\omega_l$. In the experiment, they observe the main features of the parametric amplification process, such as the exponential growth of the instability mode and subharmonic oscillations at frequency $\omega_{\rm res}/2$.

The authors also found anomalous behavior for a certain mode. The hexagonal mode $D_6$ was found to couple with the dipole motion of the condensate, leading to a dynamical instability that prevented its long-term stability. Surprisingly, similar polygonal surface modes have also been observed in classical fluids~\cite{Liu2023}. In this experiment, Liu and Wang used vertically vibrated water layers confined in parabolic containers and revealed star-shaped Faraday patterns with symmetries ranging from triangles to heptagons, closely resembling those seen in harmonically trapped Bose-Einstein condensates. In this classical setting, the hexagonal ($D_6$) mode was also found to be dynamically unstable. These observations suggest that the fragility of the hexagonal mode is not specific to quantum systems, but rather a generic consequence of complex competition among nonlinear modes.

\section{Emergent patterns from counterflow instability}
\label{sec:4}
In this section, we would like to introduce emergent pattern formation from the instabilities of hydrodynamic flows between two fluids. These counterflow instabilities have played a central role in fluid dynamics for over a century, forming the basis for phenomena such as vortex shedding, Kelvin-Helmholtz clouds~\cite{Thomson1871,Helmholtz1868}, and Rayleigh-Taylor fingers~\cite{Rayleigh1883,Taylor1950}, all of which are not only visually striking but also practically important in the fields of meteorology and aerospace engineering~\cite{Ho1984,Fritts2003}. Remarkably, these instabilities can manifest in quantum fluids, raising many intriguing questions because of the different nature of the superfluids, such as zero viscosity and different mechanisms for dissipation.  In this section, we review a series of experiments that explored the vortex array instabilities in fermionic superfluids~\cite{Hernandez2024}, Kelvin-Helmholtz instability in spinor condensates~\cite{Huh2025}, and Rayleigh-Taylor instability in a binary superfluid mixture~\cite{Geng2025}.

\subsection{The Kelvin-Helmholtz instability in spinor condensates}
\label{subsec:4.1}

\begin{figure}[t]
\centering\includegraphics[scale=.65]{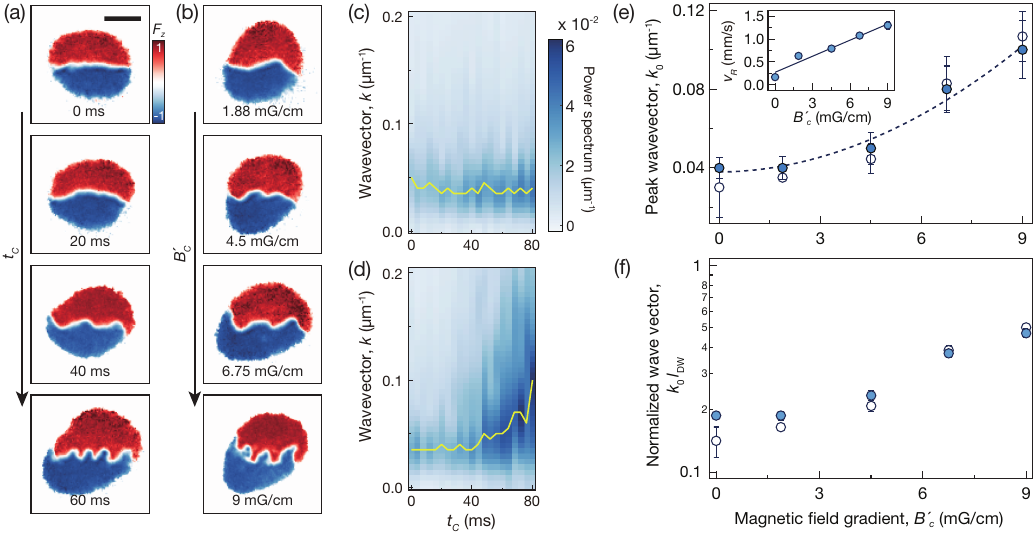}
\caption{\textbf{Kelvin-Helmholtz instability in binary superfluids.} 
(a)  Time evolution dynamics of the domain wall under a fixed gradient $B_c'=9$~mG/cm$^{-1}$.
(b)  In situ image of longitudinal magnetization density $F_z$ at a hold time $t_c=80$~ms with increasing field gradient (from top to bottom). 
Power spectra of the domain wall interface with (c) $B_c'=1.88$~mG/cm$^{-1}$ and (d) $B_c'=9$~mG/cm$^{-1}$.
(e) Peak wavenumber as a function of field gradient strength. Closed symbols mark experimental data, which show good agreement with the numerical solution of the Gross–Pitaevskii equation (open circles). Dashed line is the quadratic fit to the data. Inset depicts the linear dependence of the counterflow velocity on the field gradient. 
(f) Scaled wavenumber $k_0l_{\rm DW}$ as a function of field gradient strength. 
Adapted with permission from ref~\cite{Huh2025}, Springer Nature Limited.
}
\label{fig:KHI}  
\end{figure}

One of the most fundamental hydrodynamic instabilities of fluid flows is the Kelvin-Helmholtz instability (KHI)~\cite{Thomson1871,Helmholtz1868}. In Kelvin's seminal work~\cite{Thomson1871}, the dynamic instability of the interface between counter-flowing fluids was treated as an infinite array of vortices, a vortex sheet. Infinitesimal fluctuations of the sheet can be exponentially amplified to display a flutter finger pattern, emitting vortices and entering the turbulence regime. Although the KHI has been reported in various fluids of different length scales~\cite{Delamere2010,Smyth2012}, an ideal realization of the KHI theory is limited in classical fluids because the vortex sheet, which has a discontinuous flow across the layer, cannot become a solution of the classical hydrodynamic equation~\cite{Volovik2002}. In this regard, superfluids can be an interesting platform for testing the KHI theory. Moreover, considering the fundamental nature of superfluids, such as the quantization of vortices, it is interesting to investigate the extent to which the classical KHI theory is valid in quantum fluids. Experimental studies of the KHI were initiated by the Helsinki group using a mixture of liquid $^3$He consisting of A and B phases, observing vortices from the counterflow instability~\cite{Blaauwgeers2002,Finne2006}. However, the mechanism of vortex generation is rather different from the KHI; instead, it is Landau instability~\cite{Takeuchi2010}.

\begin{figure}[h]
\centering\includegraphics[scale=.65]{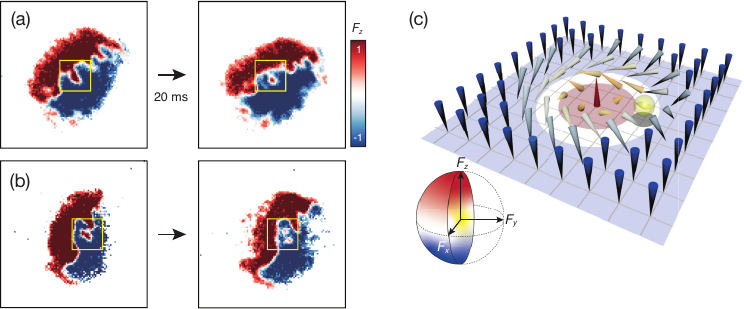}
\caption{
\textbf{Late time dynamics of the counterflow instability.}
(a) A single magnetic droplet of spin up component is emitted from the finger tip to the spin-down background magnetic domain (yellow box). 
(b) Splitting of single magnetic droplet into two separated magnetic droplets.
(c) Schematic diagram of the 2D eccentric fractional skyrmion spin texture (EFS). The EFS has half winding of the spin vector around the core. Yellow arrows represent anti-ferromagnetic spin singular point. The inset below represents the mapping of the spin vector into the spin space $\mathbf{F}$. The EFS fills half of the Bloch sphere.
Adapted with permission from ref~\cite{Huh2025}, Springer Nature Limited.
}
\label{fig:EFS}  
\end{figure}

Recently, Huh and colleagues have demonstrated the quantum counterpart of the KHI using spin-1 Bose-Einstein condensates~\cite{Huh2025}. One of the main breakthroughs in this experiment is the realization of a sharp domain wall using immiscible superfluids with a strongly ferromagnetic spinor condensate of $^7$Li atoms~\cite{Huh2020}. They induced a counterflow by applying a field gradient that exerted a force in the opposite direction.  This process effectively generates a continuous vortex sheet, where the superflow vorticity $\nabla\times\mathbf{v_s}$ with superflow velocity $\mathbf{v_s}$ is localized at the magnetic domain wall~\cite{Kokubo2021}. Fig.~\ref{fig:KHI} (a) and (b) show the time evolution of the condensate after applying the injecting counterflow. As time progressed, the domain wall became unstable and developed a flutter-finger pattern after a 40ms hold time [Fig.~\ref{fig:KHI}(a) and (b)]. By investigating the power spectrum of the domain wall interface, the exponential growth of the modulation amplitude was observed [Fig.~\ref{fig:KHI} (c) and (d)], indicating dynamic instability. Moreover, the maximal wavenumber was found to be proportional to the square of the counterflow velocity $k_0\propto v_R^2$ [Fig.~\ref{fig:KHI}(e)], which is consistent with the theoretical prediction~\cite{Takeuchi2010}. This observation is attributed to the universal feature of counterflow instability~\cite{Kokubo2021}, where KHI is always observed when the domain wall thickness is smaller than the characteristic wavelength of the interface modulation, $k_0l_{\rm DW}<1$ [Fig.~\ref{fig:KHI}(f)].

In the late-time dynamics, the flutter-finger pattern was destroyed by emitting a magnetic droplet [Fig.~\ref{fig:EFS}(a) and (b)] that could carry vorticity~\cite{Huh2025}. Similar dynamics can be observed in classical fluids, where the finger rolls up after the finger pattern fully grows~\cite{Matsuoka2014}. A detailed study showed the existence of vorticity around the magnetic core, and they found that the spin texture follows a 2D eccentric fractional skyrmion (EFS)~\cite{Takeuchi2021}. In contrast to the conventional skyrmion spin texture, the EFS has a half winding of the spin vector around its core, where the discontinuity from the spin vector can be resolved by an antiferromagnetic spin singular point~[Fig.~\ref{fig:EFS}(c)].

\subsection{Vortex array instability in fermionic superfluids}

\label{subsec:4.2}

\begin{figure}[b]
\centering\includegraphics[scale=.8]{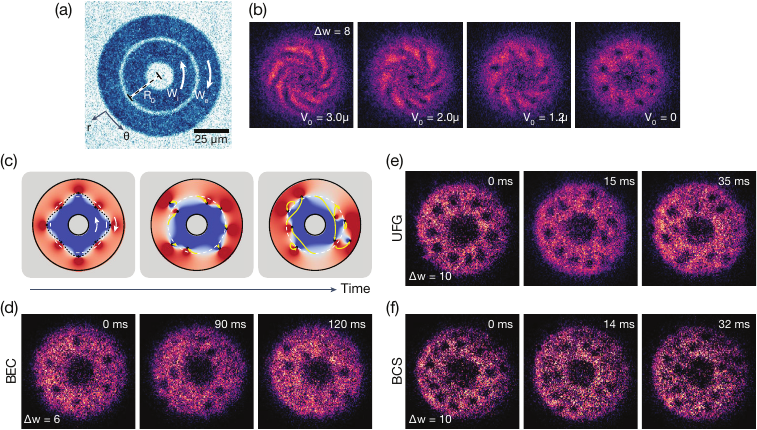}
\caption{
\textbf{Shear flow preparation and its instability in strongly interacting fermionic superfluids.}
(a) Insitu density profile of superfluid in a concentric annular trap. 
An optical barrier separates the superfluids in each region.
Counterflow was introduced by a phase imprinting technique. 
(b) Time-of-flight (TOF) images in the BEC regime at different optical barrier heights $V_0$ for phase winding $\Delta w=8$. 
At the interface between two superfluids, a vortex array with $N_v=8$ was prepared. 
(c) Numerical simulation of the vortex array dynamics.
The dotted sinusoidal line indicates the interface mode with $m=4$.
(d)-(f) Time evolution of vortex necklace (TOF image in the BEC regime) across the BEC-BCS cross over: (d) BEC side at $1/k_Fa=4.3$ with $\Delta w=6$, (e) unitary fermi gas (UFG) at $1/k_Fa=0$ with $\Delta w=10$, and (f) BCS side at $1/k_Fa=-0.3$ with $\Delta w=10$. 
Adapted with permission from~\cite{Hernandez2024}, Springer Nature Limited
}
\label{fig:array}  
\end{figure}

A similar study was conducted by Hernández-Rajkov \textit{et al}. using ultracold fermionic atoms~\cite{Hernandez2024}. They prepared two concentric annular traps, each containing a superfluid of strongly interacting $^{6}$Li atoms [Fig.~\ref{fig:array}(a)]. A mass flow in opposite directions was generated by imprinting a phase gradient onto the superfluid rings by using a far-detuned laser light~\cite{Pace2022}. As the height of the potential barrier that separates the two superfluids was lowered, they observed an array of quantized vortices [Fig.~\ref{fig:array}(b)]. Although the vortex array has a finite spatial interval so that the system cannot be described by the infinite vortex array appeared in the Kelvin-Helmholtz instability~\cite{Thomson1871}, this experiment could provide an alternative approach to study the counterflow instability in strongly interacting fermionic superfluids.

The vortex array was dynamically unstable and rearranged to form a clustering behavior~[Fig.~\ref{fig:array}(c)]. This is analogous to the growing dynamics of the flutter finger pattern in the KHI experiment using a spinor condensate~\cite{Huh2025}, where vorticity is maximized at the tip of the finger. To analyze the onset of the instability, they investigated the angular distribution of the vortices across the BCS-BEC crossover [Fig.~\ref{fig:array}(d)--(f)] and extracted the structure factor $s(m,t)=(1/N_v)\sum_{j,l}\exp[im(\theta_j(t)-\theta_l(t))]$, where $\theta_j(t)$ is the angular position of the $j$-th vortices and $m$ is a mode winding number. They observed that the growth rate of particular modes of the structure factor, for example $m=6$, grew exponentially as time evolved, $s(m,t)\propto e^{2\sigma_m t}$. Interestingly, the maximum growth rate $\sigma_m^*$ exhibits a quadratic scaling with the relative flow velocity $\Delta v$, $\sigma_m^* \propto \Delta v^2$, across the BCS-BEC crossover regime. These results imply that the scaling laws of the counterflow instability developed in classical fluids can also be observed in superfluids, even with fundamentally different natures, from weakly interacting bosonic to strongly correlated fermionic superfluids.

\subsection{The Rayleigh-Taylor instability in a binary superfluid}
\label{subsec:4.3}

\begin{figure}[t]
\centering\includegraphics[scale=.85]{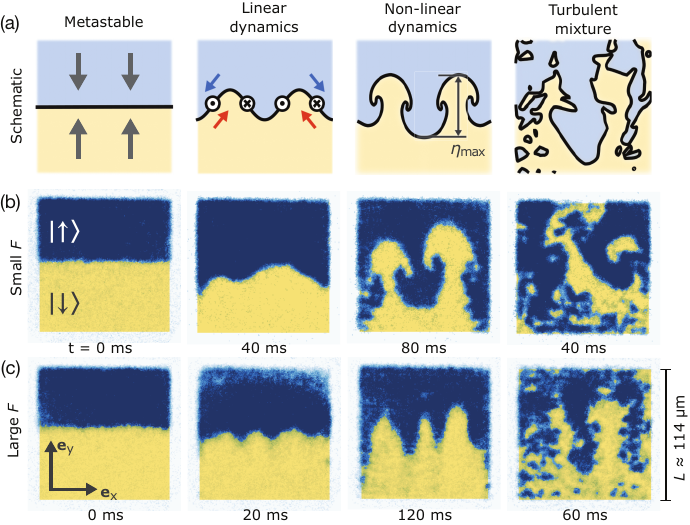}
\caption{
\textbf{Rayleigh-Taylor instability (RTI) in immiscible fluids. }
(a) Schematic diagram of the RTI. Pressing the immiscible fluids, the fluid layer became dynamically unstable to develop a wavy pattern, displaying a mushroom-shaped pattern in the nonlinear dynamic stage. 
(b) Insitu single-shot images after applying a weak $F/\hbar=-3.1(2)$~Hz/$\mu$m, (c) and strong force $F/\hbar=-15.4(8)$~Hz/$\mu$m. The interface between spin state $\ket{\uparrow}$ (blue) and $\ket{\downarrow}$ (yellow) shows the RTI. 
Adapted with permission from ref~\cite{Geng2025}, AAAS.
}
\label{fig:RTI}  
\end{figure}

The Rayleigh-Taylor instability (RTI)~\cite{Rayleigh1883,Taylor1950} occurs when a lighter fluid accelerates a heavier one under the influence of an effective gravitational field, resulting in the growth of interfacial perturbations and the eventual formation of mushroom-shaped structures,~Fig. ~\ref{fig:RTI}(a). While RTI is commonly observed in classical systems, such as inertial confinement fusion~\cite{Kilkenny1994}, oceanic mixing~\cite{Wunsch2004}, and astrophysical flows~\cite{Allen1984}, this instability has yet to be realized experimentally in quantum fluids. In particular, understanding how classical interfacial instabilities manifest in superfluid systems with quantized circulation offers a compelling opportunity to probe the correspondence between classical and quantum hydrodynamics~\cite{Sasaki2009}. 

Geng and colleagues at NIST explored the RTI using a Bose-Einstein condensate of $^{23}$Na atoms with two different hyperfine spin states, $\vert F=1,m_F=-1\rangle\equiv \ket{\uparrow}$ and $\vert F=2,m_F=-2\rangle\equiv \ket{\downarrow}$~\cite{Geng2025}. The two spin states are immiscible, and the two-component BECs were prepared in a homogeneous 2D potential with size $L=114(3)~\mu${}m. As the spin healing length $\xi_s\simeq2~\mu{}$m is much smaller than the system size $L$, this setting offers a fertile playground to observe the RTI. The dynamics was initiated by applying a field gradient, which exerted forces $F$ on the two spin states in opposite directions [Fig.~\ref{fig:RTI}]. As time progressed, the flat interface between the two condensates developed a spatial periodic modulation. Later, mushroom-[Fig.~\ref{fig:RTI}(b)] or spike-shaped~[Fig.~\ref{fig:RTI}(c)] patterns were observed, which dissolved into a turbulent mixture. The growth dynamics of the unstable mode were analyzed from the height profile $\eta_{\rm max}(t)$ [as defined in Fig. ~\ref{fig:RTI}(a)], which exhibited an exponential growth in the linear regime. 

They also provided the underlying mechanism for such instability by studying the characteristic wavenumber of the interface modulation. The dynamic instability and exponential growth dynamics can often be understood from the imaginary component of the dispersion. The dispersion of the ripplon interface mode is given by 

\begin{equation}
\omega^2=\frac{1}{2m}\left(Fk+\frac{\sigma}{\bar{\rho}}k^3\right),
\end{equation}
where $\sigma$ is the interfacial tension, and $\bar{\rho}$ is the mean density. For $F<0$, the frequency can become negative and has an imaginary component. The maximal mode gain can be found at $k_c=\sqrt{-{F\bar{\rho}}/{\sigma}}$, which shows good agreement with the experimental results. They also developed a spectroscopic technique to measure the dispersion relation by applying a periodic modulation on the force $\delta F \propto \cos(\omega_d t)$, exciting the ripplon mode with a characteristic wavenumber, and studied the dynamics of the relative velocity at the interface by relating it to the vortex number. These observations confirm that the classical description of the RTI can be extended to quantum fluids.

\newpage

\section{Outlooks}
In this review article, we have introduced representative experiments for pattern formation in superfluids of atomic gases under periodic driving. Using the state-of-the art toolkits in atomic physics, there are many new future directions to be investigated. One interesting perspective is the connection of the periodic structure in the bulk gas and the supersolidity. As recently demonstrated, the periodic density waves under periodic driving can display sound modes of supersolids~\cite{Liebster2025b}, leaving interesting questions about finite temperature effects on the crystal structure and phase coherence. This study can be further extended to pattern formation in spinor condensates. Theoretical study shows that the competition between density pattern and spin-mixing dynamics gives rise to rich structures, where in the polar phase of spin-1 condensates, a spin texture with a gas of polar core vortices and anti-vortices can emerge under periodic modulation of the scattering length~\cite{Jose2023}. Moreover, the interplay between Faraday excitation and the stripe phase in spin-orbit-coupled BECs offers a promising platform for exploring supersolid behavior with internal spin degrees of freedom~\cite{Chen2025}. As discussed above, driven quantum systems can evolve into turbulent regimes, as demonstrated in the experiments~\cite{Nguyen2019,Kwon2021,Huh2025,Geng2025}. This naturally raises questions about how ordered patterns break down, and whether certain patterns suppress or enhance the onset of turbulence. If some patterns act as a precursor to quantum turbulence, investigating how early-stage patterns seed energy cascades, vortex proliferation, and the eventual emergence of the universal Kolmogorov scaling relation could offer deeper insights into nonequilibrium dynamics of many-body quantum systems. 

On the experimental side, further development for imaging and controlling quantum gases will be required to address the above studies. 
For example, a local spin-addressing technique, developed in optical lattices~\cite{Weitenberg2011}, can be applied in spinor condensates, where a laser light can imprint an arbitrary spin pattern. The stripe phase may be generated upon modulating the laser intensity instead of modulating the scattering length between inter-spin states. Moreover, the simultaneous imaging technique in both real and momentum space can provide new insights into how patterns in the early stage would eventually transition into a turbulent regime.

\begin{acknowledgement}
The authors are thankful to generous funding from the National Research Foundation of Korea with grant numbers RS-2023-00256050, RS-2023-NR119928, RS-2025-02220735, and RS-2025-08542968.
\end{acknowledgement}

\end{document}